# Conductivity of underdoped $YBa_2Cu_3O_{7-\delta}$ : evidence for incoherent pair correlations in the pseudogap regime

B. Leridon[*], A. Défossez, J. Dumont, and J. Lesueur
*Laboratoire de Physique Quantique, ESPCI, 10, rue Vauquelin, 75005 Paris (France)*

J.P. Contour
*Unité Mixte de Physique, CNRS/Thomson-CSF, Domaine de Corbeville, 91404 Orsay cedex (France)*

Conductivity due to superconducting fluctuations studied in optimally doped $YBa_2Cu_3O_{7-\delta}$ films displays a stronger decay law in temperature than explainable by theory. A formula is proposed, which fits the data very well with two superconductive parameters, Tc and the coherence length $\xi_{c0}$, and an energy scale $\Delta^*$. This is also valid in underdoped materials and enables to describe the conductivity up to 300 K with a single-particle excitations channel in parallel with a channel whose contribution is controlled by $\xi_{C0}$, $T_c$ and $\Delta^*$. This allows to address the nature of the pseudogap in favour of incoherent pairing.

The variation of the resistivity as a function of temperature in High-Tc superconductors (HTSC) remains to elucidate and in particular why it is so different in the underdoped and in the overdoped regions. At optimal doping in $YBa_2Cu_3O_{7-\delta}$, the resistivity is linear from Tc to 325 K and extrapolates to zero resistivity at zero temperature. This cannot be explained by phonon dependency as it is well below the Debye temperature and has been depicted in the framework of the Marginal Fermi Liquid theory[1]. As soon as the material is underdoped, the resistivity exhibits no linear dependency any more but marked downward curvature from Tc to 325 K. When the material is strongly underdoped however, there seems to be a competition between this downward curvature and an upward curvature usually attributed to localisation. On the opposite, in the overdoped region of the phase diagram, the resistivity follows a rather more Fermi-liquid behaviour.

One other peculiarity of most of the underdoped cuprates lies in the density of states in the normal state, which exhibits a depression firstly discovered by NMR measurements [2, 3] and confirmed among others by specific heat [4], STM [5, 6], ARPES [7-11] measurements… The energy scale on which this depression takes place is about the superconducting gap energy, which enables to think that the pseudogap could be related to superconductivity itself. This is the hypothesis of the particle-particle channel correlations [12-16]. However other theorists attribute it to correlations in the particle-hole channel like AF correlations [17-19], charge stripes [20, 21], or quantum criticality [22-25]. The physical origin of the pseudogap is thus still highly controversial. In resistivity measurements, the above-described downward curvature is generally ascribed to the pseudogap[26]. However, since this feature is only slowly varying, one can not extract a "pseudogap temperature" T* from these measurements in lack of a quantitative model for the resistivity.

---
[*]Corresponding author: Brigitte.Leridon@espci.fr

The aim of this paper is to propose a quantitative analysis for the resistivity of under- and optimally – doped cuprates in the pseudogap regime which strongly suggests that particle-particle correlations survive well above Tc. This analysis lies upon the observation of a continuous behaviour from the fluctuation regime of the optimally doped materials to the high temperature conductivity of the underdoped ones.

The conductivity due to the superconducting fluctuations just above the critical temperature of a superconductor – usually referred to as "paraconductivity"– is strongly related to the coherence length $\xi$ and therefore to the dimensionality of the sample with respect to $\xi$. It was shown by Aslamazov and Larkin (AL)[27] that when the thickness t of the sample does not exceed $\xi$, the typical size of the superconducting fluctuating domains is limited by t and the paraconductivity is given by $\Delta\sigma_{2D}^{AL} = \left(e^2/16\hbar t\right)\varepsilon^{-1}$, where $\varepsilon = \ln(T/T_c)$ is the reduced temperature. If no dimension is smaller than $\xi$, then the fluctuating domains are three-dimensional with typical size $\xi$ and the paraconductivity is given by $\Delta\sigma_{3D}^{AL} = \left(e^2/32\hbar\xi_0\right)\varepsilon^{-1/2}$ where $\xi_0$ is the zero temperature coherence length. In 1970, Lawrence and Doniach (LD)[28] proposed a model describing lamellar superconductors, for which paraconductivity is three-dimensional close to Tc, where $\xi$ is large compared to the interlayer spacing s, and two-dimensional at higher temperatures where $\xi$ becomes smaller than s.

In high-Tc cuprates, the range over which superconducting fluctuations are important is rather large, because $\xi$ is small and the materials generally owe a strong 2D character. Therefore the study of the exact variation of the paraconductivity constitutes an alternate to evaluate $\xi_{c0}$ and to probe the dimensionality (in the LD meaning) of superconductivity in such materials. The early work of Freitas et al.[29] and Ausloos [30] showed evidence for 3D behaviour of the paraconductivity close to Tc in $YBa_2Cu_3O_{7-\delta}$ ceramics. Oh et al. [31] observed 3D/2D crossover behaviour in agreement with LD theory with $\xi_{c0}$=1.85 Å in $YBa_2Cu_3O_{7-\delta}$ thin films. It is to be noted that similar studies in $Bi_2Sr_2CaCu_2O_8$ showed 2D behaviour in the whole experimental range with no crossover or 3D tendency close to Tc [32] as expected for a more anisotropic material.

Another interesting open question was the existence of a Maki-Thompson contribution in such materials, and its temperature dependence. Such a contribution was originally proposed in low temperature superconductors by Maki [33] and Thompson [34] and was attributed to the scattering of normal-state quasiparticles by superconducting carriers. Most of the studies in HTCS to date show no such effect; furthermore, the experimentally measured fluctuation conductivity always falls below the theoretical prediction at high temperatures [29], [35], [36] and this intriguing feature has motivated the analysis presented here.

Optimally doped $YBa_2Cu_3O_{7-\delta}$ thins films of thickness about 100 nm were grown by pulsed laser deposition as described elsewhere [37]. The resistivity was measured from $T_C$ to 325 K. The paraconductivity being defined by $\Delta\sigma = 1/\rho - 1/\rho_n$, the main difficulty in such study lies in the correct evaluation of the normal state resistivity. It was deduced from a linear fit between 200 K and 270 K, the only region where $\rho_n(T)$ was strictly linear.

The data were first adjusted between $\varepsilon = 0.005$ and 0.05 using the 3D AL model for paraconductivity and $T_C = 90.3$ K, which leads to a value for the zero-temperature coherence length of $\xi_{c0}=1.84\pm0.02$ Å, consistent with the values found in the literature. The result of the fit is plotted in figure 1 (solid line). As can be seen, the adjustment is very good within the fitting range but the paraconductivity falls below the theoretical 3D AL model for $\varepsilon>0.1$.

One can remark that no Maki-Thompson term can be inferred from our measurements. Such a term would lead to an even higher discrepancy at high temperature as it would enhance the paraconductivity and modify the slope at lower temperatures leading to a poorer agreement in the range 0.01-0.1. This is consistent with the recent work of Cimberle et al. [35]. Some authors attribute this fact either to the presence of pair-breaking effects either to the d-wave character of superconductivity in such materials[36].

The high-temperature decrease of the paraconductivity could evoke a 3D/2D LD crossover but a fit with LD model does not give the correct value of s, the interlayer spacing. Furthermore, the crossover reduced temperature resulting from the fit is higher than 0.5, which means that the fit has been performed essentially over the 3D part of the LD formula. Moreover, the fit can not account for the high temperature decrease of the paraconductivity. ( See in figure 1, the dashed line). A bi-lamellar model, taking into account both the inter-plane and inter-biplane coupling, as developed by Ramallo et al.[38] was either unsuccessful to explain the data. In order to explain the high temperature behaviour, a cut-off was added in the fluctuation spectrum to the AL model ($\lambda=2\pi\xi_{c0}/u$) [39, 40](see figure 1, dotted line). This gives for the best fit $\xi_{c0} = 1.77 \pm 0.04$ Å and u = 6.2, and the quality of the fit was not better than for the LD model, and could neither explain the high temperature behaviour.

The high-temperature fluctuation regime was then analysed with a "conventional" fit for the normal state resistivity (linear fit between 200 and 270 K), and the inverse of the paraconductivity was plotted as a function of T on a log-log scale in figure 2. Surprisingly, it exhibits a linear regime between 100 K and 160 K, corresponding to $\Delta\sigma^{-1}(T) = AT^\alpha$ with $\alpha \approx 7.5 \pm 0.05$ and $A \approx 9 \times 10^{-13} \mu\Omega.cm$. This may also be written as $\Delta\sigma(\varepsilon) = \Delta\sigma_0 e^{-\varepsilon/\varepsilon_0}$ where $\varepsilon_0 = 1/\alpha \approx 0.13 \pm 0.01$ and $\Delta\sigma_0^{-1} = 420 \mu\Omega.cm$. Actually, the low temperature 3D AL behaviour as well as the high temperature behaviour may be modelled by the same simple phenomenological interpolating function $\Delta\sigma(\varepsilon) = \Delta\sigma_0 / \sqrt{2\sinh(2\varepsilon/\varepsilon_0)}$. The agreement of our data with this function is extremely good (see figure 3). Moreover, the fit performed only on the high-temperature (not AL) part of $\Delta\sigma$ gives parameters $\Delta\sigma_0$ and $\varepsilon_0$ such that the value of $\xi_{c0}$ deduced from these is exactly $\xi_{c0} = e^2 / 16\hbar\Delta\sigma_0 \sqrt{\varepsilon_0} \approx 1.8 \text{Å}$. Hence this function can be also written $\Delta\sigma(\varepsilon) = e^2 / \left[16\hbar\xi_{c0}\sqrt{2\varepsilon_0 \sinh(2\varepsilon/\varepsilon_0)}\right]$, where $\varepsilon_0$ is the only adjustable parameter and affects only the high temperature part of the curve. $\varepsilon_0$ is of the form $\ln(T^\#/Tc)$ where $T^\#$ is a characteristic temperature governing – with $\xi_{c0}$ - the collapse of the superconducting fluctuations. As $\varepsilon_0$ deduced from the fit is equal 0.13 and $T_c$ = 90.3 K, this gives $T^\#$ = 103 K.

To this point it is important to note that this high temperature behaviour can by no means be attributed to an incorrect choice for the linear normal state. If we artificially include the extra resistivity in the normal state, the value of $\Delta\sigma$ remaining for the 3D fluctuations is too low to be accounted for by AL 3D or 2D theory. Moreover, the normal state resistivity would depend on the value of Tc and would not be linear any more. We can thus attribute this high-temperature variation either to the superconducting fluctuations themselves or to a phenomenon occurring in the vicinity of Tc, affecting the fluctuation conductivity and controlled by $\xi_{c0}$, Tc and $\varepsilon_0$.

Let us now consider the case of a slightly underdoped $YBa_2Cu_3O_{7-\delta}$ thin film grown by pulsed laser deposition (Tc = 85 K). The resistivity of such material is plotted in figure 4a (open circles). As usual, the resistivity from Tc to 300 K shows a slight downward curvature. It could be fitted with an excellent agreement using a linear resistivity in parallel with the above-mentioned expression for the paraconductivity as is demonstrated in figure 4a (solid line). The fitting parameters were $\xi_{c0}$ = 0.78 ±0.01 Å, $\varepsilon_0$ = 1.34 ±0.01 and for the "linear" resistivity $\rho$ = 2.0T + 60 $\mu\Omega$.cm.

This result was extended to the data from Carrington et al. on oxygen deficient thin films[41] with an excellent agreement as well. Two examples are shown in figure 4, for $\delta$ = 0.23 (Tc=80.7 K) and $\delta$ = 0.39 ($T_c$=58.5 K). The fitting parameters are given in Table 1. However, for the more underdoped sample, whose resistivity exhibits upward curvature, a variable range hopping (VRH) term had to be added to the linear resistivity component ($\rho=\rho_0\exp(T_0/T)^{1/4}$, $\rho_0$=0.41 µΩ.cm and $T_0$=$10^5$K). It is to be noted that the sensitivity of the fits to the value of $\xi_{c0}$ and $T_c$ was very high, so the resistivity had to be determined with great care in order to extract the correct $\xi_{c0}$.

From these results three consequences can be outlined:

- Something happens around 100 K in optimally doped $YBa_2Cu_3O_{7-\delta}$, which strongly modifies and lowers the AL superconducting fluctuations. This phenomenon is controlled by the parameters $T_c$, $\xi_{c0}$ and $\varepsilon_0$. This happens also in underdoped $YBa_2Cu_3O_{7-\delta}$ thin films, and the resistivity may still be described, until room temperature, by the same interpolating function with a corresponding decrease of $\xi_{c0}$[†] and increase of $\varepsilon_0$, as $T_c$ is lowered. We therefore clearly establish a link between the superconducting properties and the conductivity in the pseudogap regime. This strongly supports the particle-particle correlation channel for the pseudogap.

- The fact that the transition can still be very well described by an AL model is of great importance: it means that the correlations existing at high temperature (preformed pairs[12] or incoherent pair excitations [14, 15, 42]) do not alter the system susceptibility with respect to the OP amplitude fluctuations in the vicinity of $T_c$. In particular it could mean that these correlations *do not transform into* pairs but coexist with AL fluctuating pairs above $T_c$ and maybe with coherent pairs below $T_c$. Moreover, this possibly rules out particle-particle correlations models based on a fluctuating phase of the OP since in the presence of preformed pairs one should not expect AL fluctuations of the order parameter near $T_c$. This might be in favour of a Bose-Einstein-Condensation-BCS crossover scenario[42], where three different types of "objects" coexist below $T_c$: superconducting pairs, incoherent pairs related to the pseudogap and quasiparticles. The incoherent pairs would be responsible for the high temperature lowering of the resistivity in the pseudogap region, and the partial suppression of AL fluctuations since they compete with superconductivity[15].

- Considering that we are now able to extract the "superconductive" part of the conductivity of the "normal" state it is possible to draw conclusions about the single-particle excitations part. The resistivity of such excitations is still linear in underdoped $YBa_2Cu_3O_{7-\delta}$. For more underdoped samples, it is gradually replaced by an upward variation attributable to VRH transport.

---

[†] This variation for $\xi_{c0}$ is in accordance to what is reported for underdoped materials in the litterature

A natural question arises: What is the meaning of $\varepsilon_0 = Ln(T^\#/Tc)$ ? Or is $T^\#$ equal to $T^*$? Actually, there is no drastic change neither at $T^*$, nor at $T^\#$ in resistivity of $YBa_2Cu_3O_{7-\delta}$. The fit is not very sensitive to a small variation of $T^\#$, but all previous determination of $T^*$ from resistivity measurements suffers from the same uncertainty [26]. It appears here that $\varepsilon_0$ is more to be related to an energy scale $\Delta^*$ of about 9-50 meV (See Table 1). The orders of magnitude of $\Delta^*$ compare nicely to the values of the pseudogap energy calculated by Chen et al.[15] in the underdoped region, or to other experimental values (in $Bi_2Sr_2CaCu_2O_8$ )[6].

Another important address of this work is the inner resistivity of underdoped materials. In the absence of pairing correlations, the resistivity is to be linear or VRH-like. This constitutes an alternate to the experiments of Ando *et al*.[43-45] on $Bi_2Sr_2CaCu_2O_8$ and $La_{2-x}Sr_xCuO_4$. When superconductivity is destroyed by a pulsed magnetic field of 60 T, the normal state still exhibits upward curvature at low temperature. It would be of great interest to do the experiment at higher temperatures.

Study of the high-temperature part of the conductivity due to fluctuations in optimally doped $YBa_2Cu_3O_{7-\delta}$ sheds a new light on the resistivity of underdoped cuprates. There seems to be continuity between the behaviour of AL paraconductivity in the optimally doped material, which is affected at higher temperatures by a phenomenon governed by Tc, $\varepsilon_0$ and $\xi_{c0}$, and the conductivity of weakly underdoped material. The latter can be modelled with two components, one varying linearly, presumably due to the single-particle excitations, and one which could be describing the particle-particle correlations (with three parameters Tc, $\xi_{c0}$ and $\varepsilon_0$). This function evolves naturally into the AL paraconductivity at low temperature with the same value for $\xi_{c0}$.

The analysis developed in the present paper is consistent with the coexistence of pairing correlations and single-particle excitations such as developed for example by Chen et al.[15] but not with models for the pseudogap based on particle-hole correlations, as there would not be physical reasons for them to be related to $\xi_{c0}$ and Tc. The presence of AL fluctuations (related to the amplitude of the order parameter) is also more in favour of a three fluids model like the BCS-BEC crossover than a fluctuating phase model.

We thank M. Aprili for useful discussions. This work was supported by grants from the French Ministère de la Recherche (AD and JD).

**Figure Caption:**

**Figure 1:**

Paraconductivity as a function of reduced temperature for optimally doped $YBa_2Cu_3O_{7-\delta}$ thin film (Log-log scale). Open circles: experimental data ; solid line : Aslamazov-Larkin 3D model with $\xi_{c0}$ =1.84 Å ; dotted line : Aslamazov-Larkin 3D with cut-off frequency ; dashed line : Lawrence-Doniach model with interspacing s=5±1Å et $\xi_{c0}$ =1.82 Å.

**Figure 2:**

Inverse of the paraconductivity of an optimally doped $YBa_2Cu_3O_{7-\delta}$ thin film as a function of temperature (log-log scale) exhibiting linear behaviour from 100 K to 160 K.

**Figure 3:**

Fit of the paraconductivity as a function of $\varepsilon$ with the interpolating formula $\Delta\sigma(\varepsilon) = e^2 / \left[ 16\hbar\xi_{c0}\sqrt{2\varepsilon_0}\sinh(2\varepsilon/\varepsilon_0) \right]$. The fitting parameters are $\xi_{c0}$ =1.84 Å, $\varepsilon_0$ = 0.13, Tc=90.3 K.

**Figure 4 :**

Open circles : resistivity as a function of T for an underdoped $YBa_2Cu_3O_{7-\delta}$ thin films. Solid line : interpolating function with parameters $\xi_{c0}$ and $\varepsilon_0$ . (a) Underdoped film (Tc=85K), $\xi_{c0}$ = 0.78 Å, $\varepsilon_0$ = 1.34. (b) Underdoped film with $\delta$ = 0.23 from Carrington *et al.*[41], $\xi_{c0}$ = 0.62 Å and $\varepsilon_0$ = 2.08. (c) *ibid.* with $\delta$=0.39 $\xi_{c0}$ = 0.42 Å and $\varepsilon_0$ = 1.97.

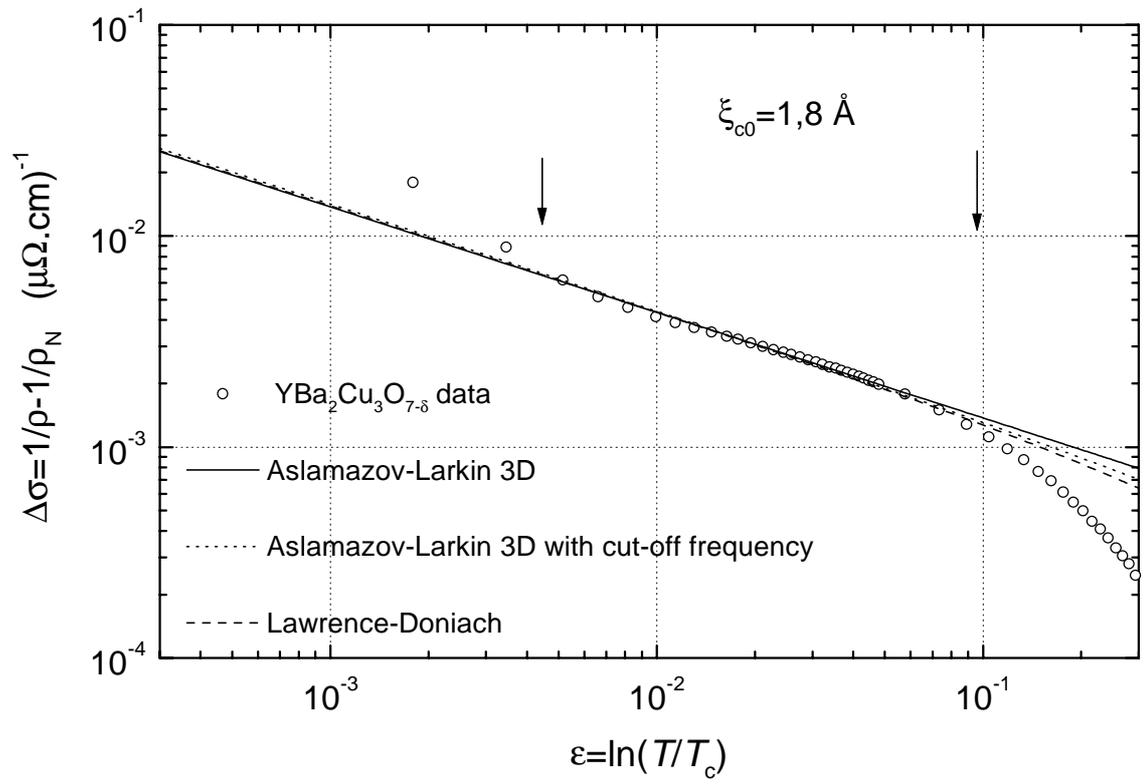

Figure 1

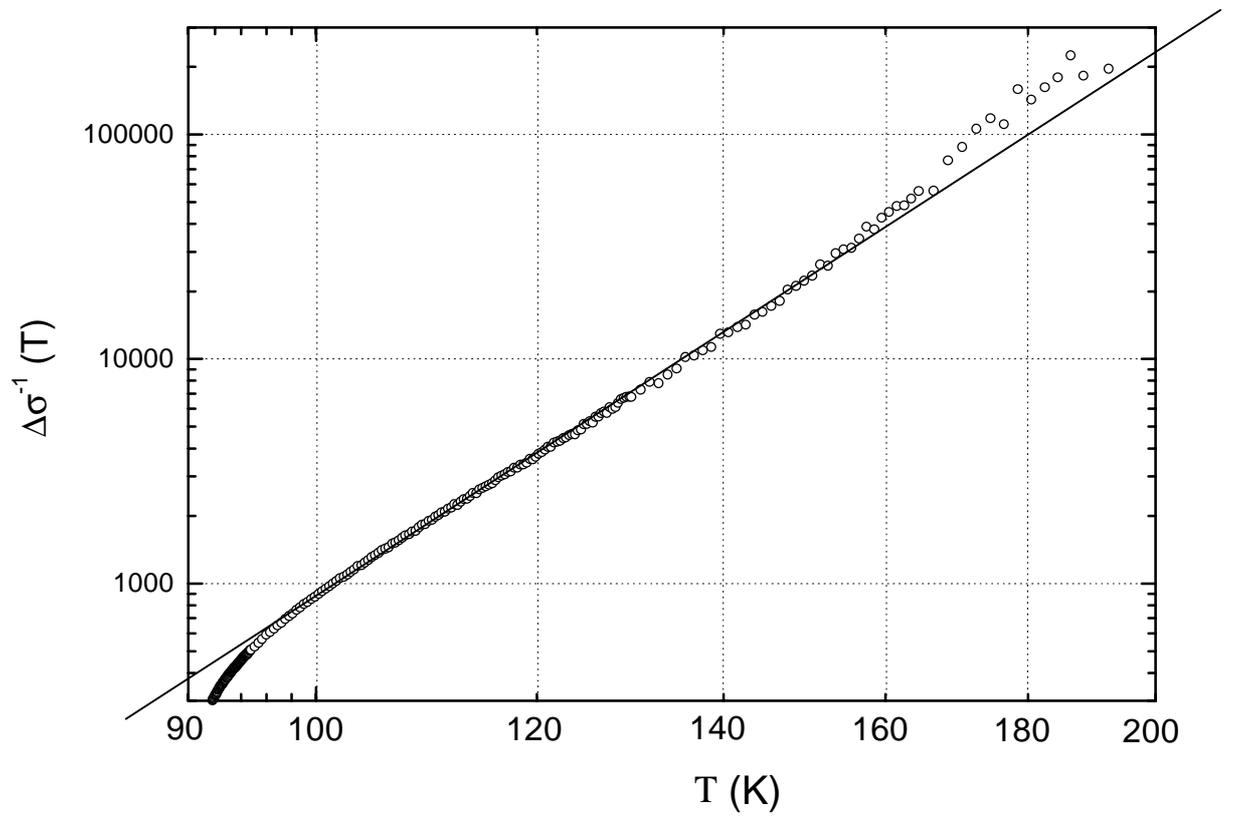

Figure 2

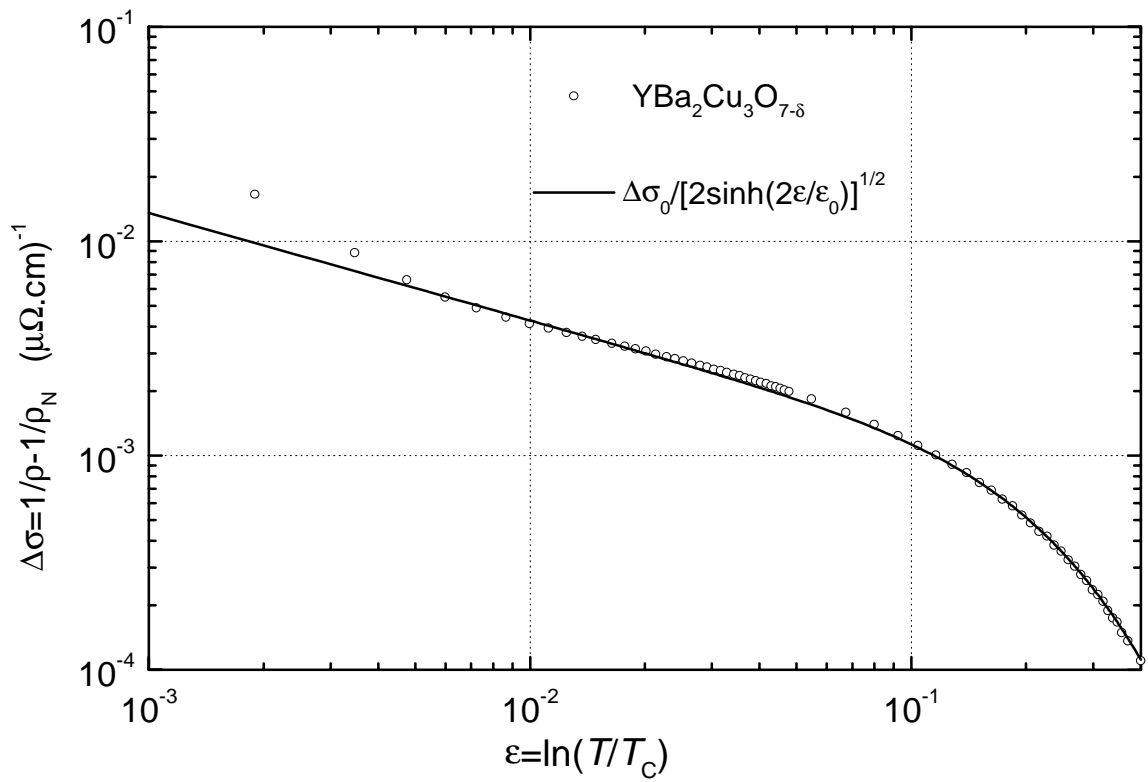

Figure 3

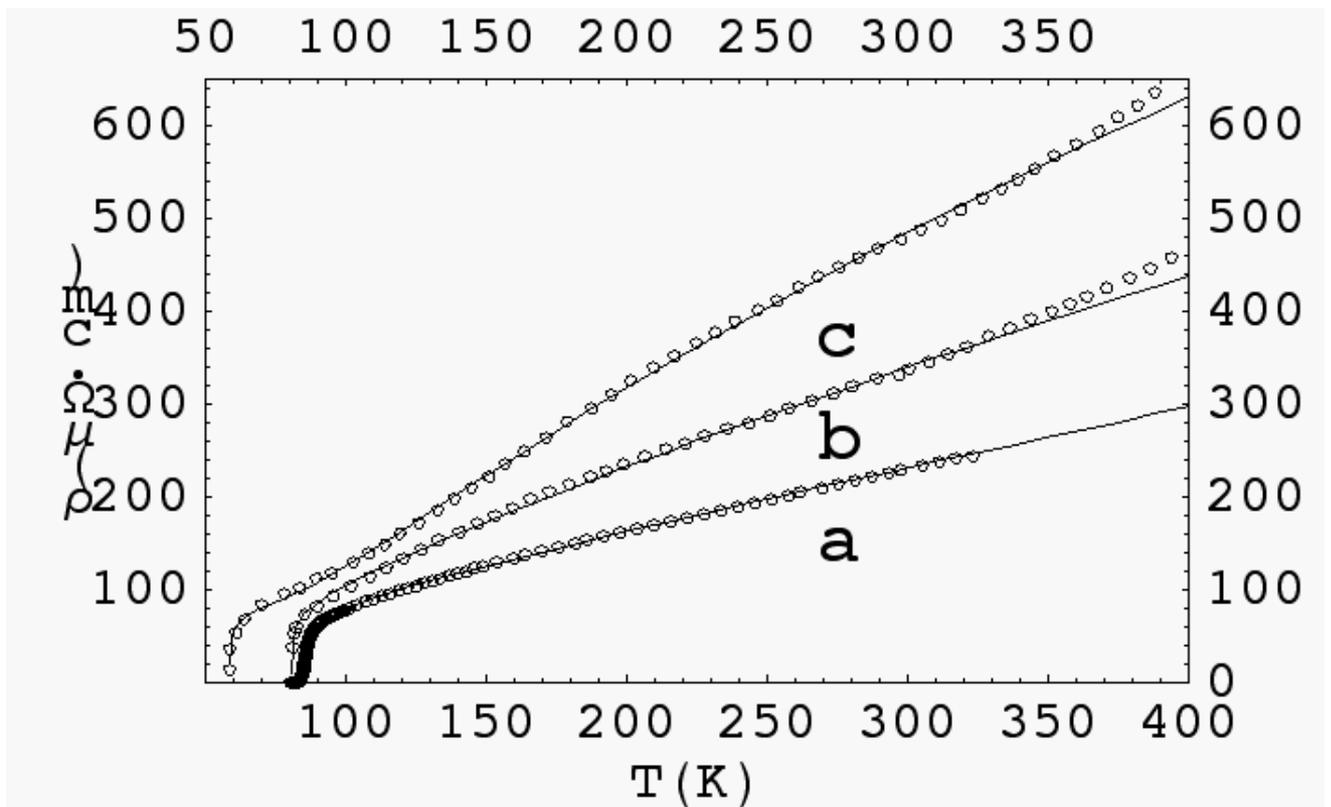

Figure 4

**Table Caption :**

**Table 1 :** Fitting parameters and typical energy $\Delta^*$ for four different samples. The last two ones are from Carrington et al.[41].

**Table 1 :**

| Sample | Tc (K) | $\xi_{C0}$ (Å) ±0.02 | $\epsilon_0$ ±0.02 | $\Delta^*$ (meV) |
|---|---|---|---|---|
| Opt.doped $YBa_2Cu_3O_{7-\delta}$ thin film | 90.3 | 1.84 | 0.13 | 9 |
| Underdoped $YBa_2Cu_3O_{7-\delta}$ thin film | 85 | 0.78 | 1.34 | 28 |
| $YBa_2Cu_3O_{7-\delta}$ thin films ($\delta$=0.23)[41] | 80.7 | 0.62 | 2.08 | 51 |
| $YBa_2Cu_3O_{7-\delta}$ thin films ($\delta$=0.39)[41] | 58.5 | 0.42 | 1.97 | 36 |